\documentclass[12pt]{article} 

\usepackage{scicite}

\usepackage{times}

\usepackage{amsmath}
\usepackage{amsfonts}
\usepackage{amssymb}
\usepackage{graphicx}

\usepackage{booktabs}

\newenvironment{sciabstract}{%
\begin{quote} \bf}
{\end{quote}}

\usepackage[labelfont=bf,figurename=Fig.,labelsep=period,font=footnotesize]{caption}

\title{\Large\bf Group mixing drives inequality in face-to-face gatherings}

\author
{\doublespacing 
M. Oliveira,${}^{1\ast}$ F. Karimi,${}^{1,2}$ M. Zens,${}^{1}$ J. Schaible,${}^{1,4}$  M. G\'enois,${}^{3}$ M. Strohmaier${}^{5,1,2}$\vspace{.1in}\\
\footnotesize{${}^{1}$Computational Social Science, GESIS--Leibniz Institute for the Social Sciences, Germany}\\
\footnotesize{${}^{2}$Complexity Science Hub Vienna, Josefst\"{a}dterstra{\ss}e 39, 1080 Vienna, Austria}\\
\footnotesize{${}^{3}$Aix Marseille Univ, Universit\'{e} de Toulon, CNRS, CPT, Marseille, France}\\
\footnotesize{${}^{4}$TH-K\"oln - University of Applied Sciences, Gummersbach, Germany}\\
\footnotesize{${}^{5}$Business School, University of Mannheim, Germany}\\
\footnotesize{$^\ast$To whom correspondence should be addressed; E-mail:  moliveira@tuta.io.}\vspace{-.5cm}
}

\date{}

\usepackage{changepage}

\usepackage{xcolor}
\usepackage{soul}
\definecolor{reddish}{HTML}{FBB4AE}
\definecolor{blueish}{HTML}{B3CDE3}
\definecolor{magentish}{HTML}{FF00AA}
\definecolor{greenish}{HTML}{a1d99b}

\usepackage{hyperref}

\usepackage{setspace}

\usepackage{lineno}

\begin{document} 




\maketitle 




\begin{sciabstract}
Uncovering how inequality emerges from human interaction is imperative for just societies. Here we show that the way social groups interact in face-to-face situations can enable the emergence of disparities in the visibility of social groups. These disparities translate into members of specific social groups having fewer social ties than the average (i.e., degree inequality). We characterize group degree inequality in sensor-based data sets and present a mechanism that explains these disparities as the result of group mixing and group-size imbalance. We investigate how group sizes affect this inequality, thereby uncovering the critical size and mixing conditions in that a critical minority group emerges. If a minority group is larger than this critical size, it can be a well-connected, cohesive group; if it is smaller, minority cohesion widens degree inequality. Finally, we expose the under-representation of individuals in degree rankings due to mixing dynamics and propose a way to reduce such biases. 
\end{sciabstract}

\baselineskip24pt 

\section*{Introduction}
Face-to-face interaction is a fundamental human behavior, shaping how people build and maintain social groups by segregating themselves from others\cite{goffman1967interaction,Kendon1975,Duncan1977,Bargiela-Chiappini2009}. This segregation can generate or exacerbate intergroup inequality in networks, producing unequal opportunities in different aspects of people's lives, such as education, employment, and health\cite{Finneran2003,Calvo-Armengol2004,Curran2005,DiMaggio2011,Stadtfeld2019}, especially when individuals tend to interact with similar others\cite{DiMaggio2012}. Though crucial to a just society, however, our understanding of the interplay between group dynamics and the emergence of inequalities in social gatherings remains limited and quantitatively unexplored.


With deep roots in sociology and anthropology\cite{goffman1967interaction,Kendon1975,Duncan1977}, the study of face-to-face interaction has advanced considerably in recent years, mainly because of new tracking devices providing fine-grained data on human interaction\cite{Fournet2014,Watanabe2014,Barrat2015,genois2019building,SchaibleHandbook}. This data has enabled researchers to uncover several properties in the way people interact with others\cite{Hui2005,Salathe2010,Cattuto2010,Stehle2011,Takaguchi2011,Isella2011a}. For example, in a social gathering, individuals tend to interact with an average number of individuals, a quantity that depends on the social occasion\cite{Isella2011a}. Though this number exhibits a trend across individuals in a gathering, the duration of each interaction lacks a central tendency---it might last from a few seconds to a couple of hours\cite{Salathe2010,Cattuto2010,Stehle2011}. These two properties have been found universally in many distinct social situations, such as schools and workplaces, indicating the existence of fundamental mechanisms underlying face-to-face interaction.


Simple social mechanisms can explain these properties as a result of local-level decisions based on individuals' attributes\cite{Zhao2011b,Starnini2013,Zhang2016,Flores2018}. One crucial attribute in face-to-face situations is the so-called \textit{attractiveness}. People with high attractiveness are more likely to stimulate interaction with others. This principle, together with individuals' social activity, constitutes a mechanism that explains well the properties found in empirical data\cite{Starnini2013} and has been extended to describe other features in face-to-face interaction, such as people's tendency to engage in recurrent interaction\cite{Zhao2011b,Flores2018}. However, focusing solely on individuals' attractiveness neglects the crucial role of pairwise interactions and social groups. For example, someone with a high intrinsic attractiveness might feel unwelcome in a community if this individual does not belong to that particular social group or fail to share the same social traits. In this paper, we refer to social groups as a group of people who share similar characteristics~\cite{tajfel2010social}. 

How social groups interact can define the position of individuals in networks, particularly when groups have unequal sizes. For instance, the smallest group (i.e., the minority group) in a network can have a systemic disadvantage of being less connected than larger groups, depending on the group mixing\cite{Karimi2018}. Having a lower number of connections poses several disadvantages to individuals, such as low social capital\cite{McDonald2011}, health  issues\cite{Haas2010}, and perception biases\cite{Lee2019}. Yet, the mechanisms underlying group dynamics and their relation to degree inequality in social gatherings are still unexplored. 


Here we show numerically, empirically, and analytically that degree inequality in face-to-face situations can emerge from group imbalance and group mixing. We present a mechanism---the attractiveness--mixing model---that integrates mixing dynamics (i.e., pairwise preferences) with individual preferences, expanding the established attractiveness paradigm. While attractiveness is an intrinsic quality of the individual, mixing dynamics manifest between pairs of individuals; together, they form what we call \emph{social attractiveness}. The mechanism reproduces the intergroup degree inequality found in six distinct data sets of face-to-face gatherings. With the analytical derivation of our model, we further demonstrate the impact of group imbalance on degree inequality, finding a critical minority group size that changes the system \emph{qualitatively}. When the minority group is smaller than this critical value, higher cohesion among its members leads to higher inequality. Finally, we expose the underrepresentation of minorities in degree rankings and propose a straightforward method to reduce bias in rankings. 

Previous research has overlooked group degree inequality in face-to-face gatherings, proposing that primarily the individuals' \emph{intrinsic} attractiveness controls their connectivity. We demonstrate, however, that when we ignore pairwise interaction (i.e., group mixing), we are unable to explain inequalities in degree ranking and mixing patterns as observed in empirical data. We note that our model is distinct from network models, such as stochastic block models or Barab\'{a}si--Albert model, in that we consider the spatiotemporal constraints in social gatherings, which leads to the emergence of fundamental properties in face-to-face interaction. Furthermore, our analytical contribution enables us to investigate regimes of degree inequality due to group mixing. With this work, we show how and when group imbalance leads to inequality in social gatherings, building new opportunities to understand and alleviate inequality in society.

\section*{Results}

\begin{table}[b!]
\caption{\textbf{The social networks from six different studies.} The number of nodes~$N$, the minority fraction~$f_0$, the number of edges~$E$, and the average degree of the minority $\langle k_0\rangle$, majority $\langle k_1\rangle$, and overall $\langle k\rangle$. \label{tabledata}}
\centering
\begin{tabular}{lrrrrrr}
\toprule
Data set     & \multicolumn{1}{c}{$N$} & \multicolumn{1}{c}{$f_0$} & \multicolumn{1}{c}{$E$} & \multicolumn{1}{c}{$\langle k_0\rangle$} & \multicolumn{1}{c}{$\langle k_1\rangle$}  & \multicolumn{1}{c}{$\langle k\rangle$}   \\\midrule

School 1 &  $242$ & $0.46$ & $8,317$ & $64.85$ & $72.08$  & $68.74$  \\
School 2 &  $126$ & $0.33$ & $1,710$ & $24.71$ & $28.32$  & $27.14$  \\
School 3 &  $180$ & $0.27$ & $2,220$ & $23.46$ & $25.11$  & $24.67$  \\
School 4 &  $327$ & $0.44$ & $5,818$ & $36.48$ & $34.87$  & $35.58$  \\
Conference 1 &  $115$ & $0.43$ & $5,508$ & $95.35$ & $96.12$  & $95.79$  \\
Conference 2 &  $202$ & $0.30$ & $11,412$ & $118.61$ & $110.56$  & $112.99$  \\
\bottomrule 
\end{tabular}
\end{table}

We study the dynamics of social gatherings using six different data sets\cite{SocioPatterns}, four from schools\cite{Stehle2011,Fournet2014,Mastrandrea2015} and two from academic conferences\cite{genois2019building}. Each data set consists of the individuals' interactions captured via close-range proximity directional sensors that individuals wore during each gathering. With these data sets, we construct the social network of each gathering, where a node is an individual and an edge exists if two individuals have been in contact (i.e., face-to-face interaction) at least once during the study (Table~\ref{tabledata}). In these networks, the degree distributions exhibit a single peak at the center (see Section 1.1 in Supplementary Material).
The data further contain gender information on individuals, which allows us to define two groups in each network. Thus, the notion of group refers to individuals who share a similar characteristic with regards to gender. In our manuscript, we refer to group mixing as the systematic preference of group members to interact with individuals from specific social groups (including their same group). 

In all considered data sets, there were fewer female participants than male participants. Throughout this article, we refer to the smallest group as the minority group, and we denote $0$ as the label for the minority group and $1$ for the majority.

\begin{figure}[b!]
\centering
\includegraphics[width=6in]{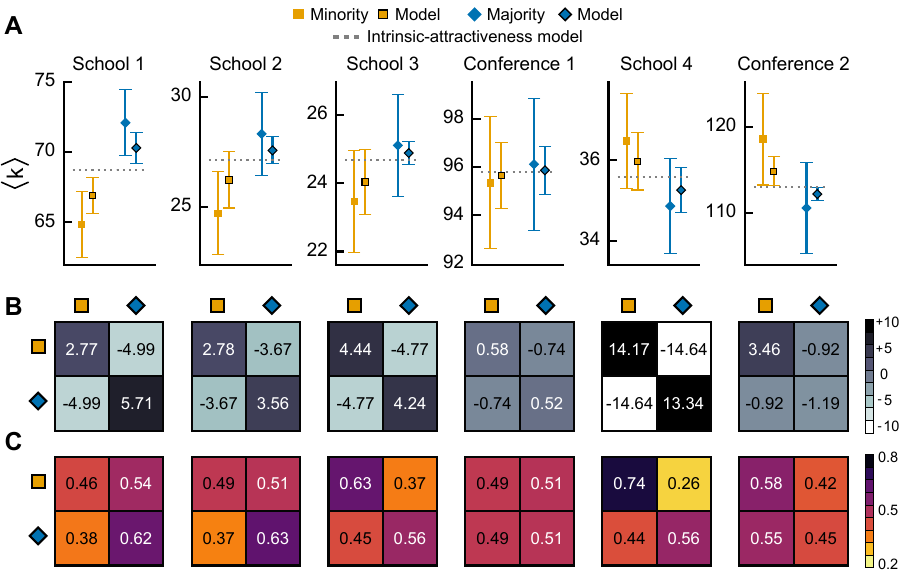}
\caption{\textbf{Face-to-face interaction: degree inequality and group mixing.} 
(\textbf{A}) The average degree of the minority and majority groups in six empirical data sets of face-to-face interaction. Systematically, some groups have a higher average degree than others. An intrinsic-attractiveness model (dashed lines) is unable to explain these differences. The error bars are the standard error of the mean. (\textbf{B}) The intra- and inter-group interaction in these social gatherings. The $z$ score values correspond to the comparison against the null model: the higher the value, the more often groups interact. In most cases, individuals are more likely to interact with individuals from the same group. (\textbf{C}) The estimated mixing matrix $\mathbf{H}$ from the data sets. The matrix describes the likelihood of groups connecting among themselves. With the estimated mixing matrix, we simulate the attractiveness--mixing model and (\textbf{A}) show that the model reproduces the average degree inequality observed in the data.
}
\label{fig:fig1}
\end{figure}

\subsection*{Degree inequality and mixing in face-to-face interaction}

To characterize the connectivity of the groups in the networks, we measure the average degree of individuals in each group, finding a systematic degree inequality among groups (Fig.~\ref{fig:fig1}A). The minority group exhibits a lower average degree than the majority group in School~1, School~2, and School~3, whereas the opposite occurs in Conference~2 and School~4, and both groups have the same average degree in Conference~1. Though this degree inequality arises in face-to-face situations, an intrinsic-attractiveness model of face-to-face interaction \cite{Starnini2013} fails to explain the group differences (dashed line in Fig.~\ref{fig:fig1}A), since it neglects group mixing in social gatherings.

To understand how groups mix in the networks, we examine the inter- and intragroup ties by comparing them with the configuration model (see Methods). We find that individuals were more likely to interact with individuals from the same group, indicating that homophily\cite{mcpherson1987homophily} plays a significant role in face-to-face situations. In most of the networks, our results show that intragroup ties are more frequent than what one would expect by chance (Fig.~\ref{fig:fig1}B).

However, though fundamental in shaping networks, this group mixing cannot emerge in the intrinsic-attractiveness paradigm because it lacks relational attributes. In this paradigm, systematic variations in individuals' attributes can lead to group differences, but they fail to form mixing patterns as observed in data. For example, consider a social gathering in which the members of the minority group have lower intrinsic attractiveness than the majority group. In this scenario, according to the intrinsic-attractiveness paradigm, the minority group would tend to interact with the majority group, boosting the majority group average degree. This setting would explain group degree inequality. However, in this same setting, the members of the minority group would be less prone to interact with their own group---the opposite of what occurs in the data, where intragroup mixing is significant (Fig.~\ref{fig:fig1}B; see also Section 1.2 in Supplementary Material). In other words, variations in individuals' attractiveness are insufficient to explain the degree inequality observed in reality. To uncover the underlying mechanism of these social dynamics, we need to disentangle the individuals' intrinsic attractiveness and the relational attributes in face-to-face interaction.

\subsection*{Modeling mixing in social interaction}
\begin{figure}[b!]
\begin{adjustwidth}{-.5in}{0in} 
\includegraphics[width=7.2in]{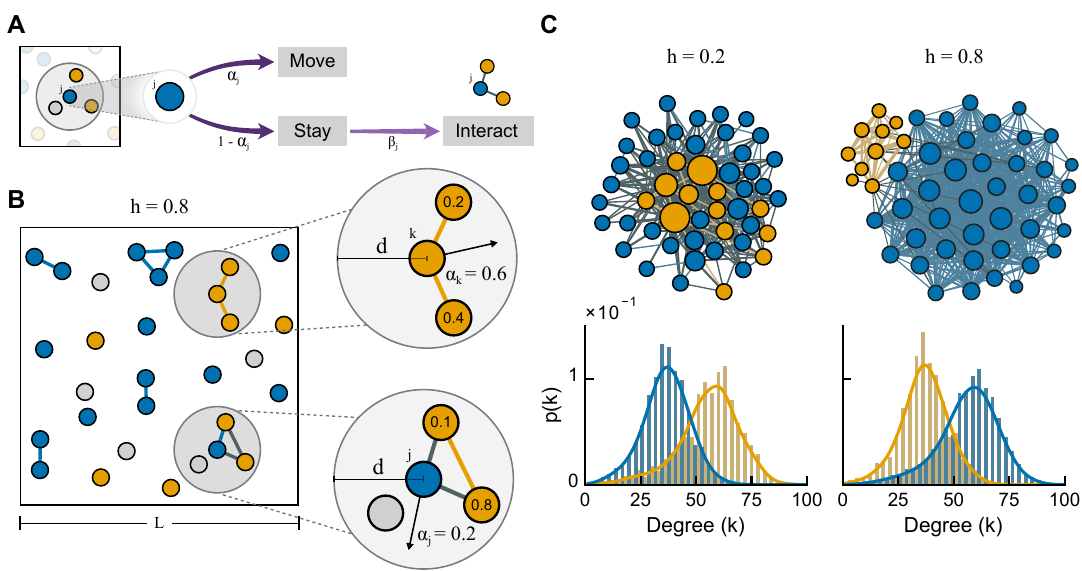}
\end{adjustwidth}
\caption{\textbf{The attractiveness--mixing model: the micro- and macro-level behavior.} 
(\textbf{A}) A schematic description of the micro-level interaction among individuals in the attractiveness--mixing model. Each individual moves across space and forms social ties based on their group membership and the composition of their vicinity. \textbf{(B)}~In the figure, nodes represent individuals, and the labels are their intrinsic attractiveness; nodes' color represents group membership, where gray indicates inactive nodes. In this example, the probability of same-group mixing is $h = 0.8$. At this time step $t$, individual $k$ moves with probability $\alpha_k(t) = 0.6$ because of their vicinity; with the complementary probability, this individual does not move and interacts with their neighbors with probability $\beta_k(t)=0.8$. Similarly, individual $j$ moves with probability $\alpha_j(t)=0.2$; otherwise this individual  stays and interacts with probability $\beta_j(t)=0.2$. \textbf{(C)} Two examples of the model's macro-level behavior. When the probability of same-group mixing is low ($h=0.2$), the minority group tends to connect to the majority group, and vice versa. In this scenario, group imbalance leads minority members to interact with more individuals (i.e., higher degree) than the majority members. High same-group mixing probability ($h=0.8$) leads minority members to attract individuals from the minority group. In this case, majority members have a higher degree centrality than the minority group. }
\label{fig:fig2}
\end{figure}

We present the \textit{attractiveness--mixing} model that incorporates (i) intrinsic attractiveness of individuals and (ii) relational attributes between groups. We show that these ingredients are sufficient to explain the degree inequality observed  in social dynamics with minority groups. In the model, each individual has an intrinsic attractiveness and belongs to a group. The members of a group share the same mixing tendency, which regulates the pairwise interactions. In general, individuals move across space and form social ties depending on their group membership and the composition of their surroundings (see Fig.~\ref{fig:fig2}A-B). 

In the attractiveness--mixing model, an individual $i$ has three attributes: a group label $b_i \in [0, B - 1]$, where $B$ is the number of groups; an intrinsic attractiveness $\eta_i \in [0, 1]$; and an activation probability $r_i \in [0, 1]$.
The mixing patterns in this system are encoded in the $B\times B$ mixing matrix $\mathbf{H}$. Each row of $\mathbf{H}$ can be seen as a probability mass function that weighs the likelihood of group interaction. 
In this model, $N$ individuals perform random walks in a two-dimensional $L\times L$ periodic space and move based on the composition of their vicinity. We define $\mathcal{N}_i(t)$ as the set of individuals who are within radius $d$ of individual $i$ at time $t$. We denote $n_b$ as the size of a group $b$ and $f_b=n_b/N$ as the group fraction. The individuals move only probabilistically. At each time step $t$, each individual $i$ moves with probability
\begin{equation}
    \alpha_i(t) = 1 - \max_{j \in \mathcal{N}_i(t)}\{\eta_j\}
    \label{eq:alpha}
\end{equation}
and a step of constant length $v$ along a random direction of angle $\xi \in (0, 2\pi]$. With the complementary probability, individual $i$ does not move and has the chance to interact with individuals in the vicinity depending on the group mixing likelihood. Precisely, individual $i$ interacts with their neighbors of highest mixing likelihood with probability
\begin{equation}
    \beta_i(t) = \max_{j \in \mathcal{N}_i(t)}\{h_{b_ib_j}\},
    \label{eq:beta}
\end{equation}
where $h_{b_ib_j}$ is an element of $\mathbf{H}$ and denotes the mixing probability between $b_i$ and $b_j$ (see Section 3 in Supplementary Material for pseudocode). Overall, an individual interacts with other individuals depending on their intrinsic attractiveness and social group mixing; together these two ingredients form the \textit{social} attractiveness. Finally, individuals can be active or inactive; they only move and interact with others if they are active. An inactive individual $i$ becomes active with probability $r_i$, whereas an active but isolated individual $i$ becomes inactive with probability $1-r_i$. In this study, we assume that the intrinsic attractiveness $\eta_i$ and the activation probability $r_i$ come from a continuous uniform distribution in $[0, 1]$.

The mixing matrix $\mathbf{H}$ and the group sizes have a significant impact on the model dynamics, affecting individuals' connectivity, especially when groups have different sizes. For example, in a system having two groups, a minority group with proportional size $f_0=0.2$ and a majority group with $f_1=0.8$, the mixing dynamics lead the system to be in different regimes that influence average group degree (Fig.~\ref{fig:fig2}C). When intragroup interaction is less likely than intergroup interaction (i.e., $h < 0.5$), the system is in a \emph{heterophilic} regime, and the minority group has a higher degree than the majority. This degree disparity arises because of the majority group favoring interaction with a small number of people (i.e., the minority group), thereby reducing the majority group connectivity. The opposite occurs in a \emph{homophilic} regime, where intragroup interaction is more likely than intergroup interaction (i.e., $h > 0.5$), which results in the minority having a low average degree.   

To investigate group degree inequality, we derive the model analytically to uncover the impact of mixing dynamics and group sizes on inter- and intragroup edges. Without loss of generality, we focus on the case of two groups, $B=2$, finding the closed-form expressions for the normalized group edge matrix, $e_{rs} = E_{rs}/E$, and the mixing matrix $\mathbf{H}$ (see Methods for details). The normalized intragroup edges are given by
\begin{equation}
\label{eq:e00}
    e_{00} = \dfrac{f_0^2 (1-h_{01}^2) }{f_0^2 (1-h_{01}^2)  + 2f_0 f_1 (1-h_{00}h_{11})  + f_1^2(1-h_{10}^2)}
\end{equation}
and
\begin{equation}
\label{eq:e11}
    e_{11} = \dfrac{f_1^2 (1-h_{10}^2) }{f_0^2 (1-h_{01}^2)  + 2f_0 f_1 (1-h_{00}h_{11})  + f_1^2(1-h_{10}^2)},
\end{equation}
and similar expressions exist for the intergroup edges (see Methods). 
With these expressions, not only can one study the model dynamics, but one can also estimate the mixing matrix from empirical networks and assess the model's ability to explain data. 

\subsubsection*{Estimating the mixing matrix from data}
To compare the attractiveness--mixing model with the empirical data, we simulate the model with the parameters as estimated from the data. Our results show that the model reproduces the average degree inequality observed in the networks (Fig.~\ref{fig:fig1}A) by comparing the distributions via  Kolmogorov--Smirnov test (see Table 3 in Supplementary Material). In addition, this model reproduces other properties found in face-to-face gatherings such as the distributions of interaction duration, inter-interaction time and weight distribution (Fig.~8 in Supplementary Material). We use Eq.~(\ref{eq:e00}) and Eq.~(\ref{eq:e11}) to estimate the mixing matrix $\mathbf{H}$ from the data (Fig.~\ref{fig:fig1}C). This matrix tells us the tendency of groups to interact among themselves in each face-to-face opportunity. With the estimated matrices, we simulate the model using the same number of nodes and group sizes in each data set. We find that mixing dynamics and group imbalance enable the emergence of the group disparities observed in the data. 

We highlight that the degree inequality can favor the minority or majority group. For example, in the School~4 and Conference~2 data sets, the smallest group tends to have more connections than the largest group. The model exhibits the same tendency. From a model perspective, this phenomenon can occur because of the asymmetry in the group mixing. To understand such cases better, we delve into the model and its regimes. 

\subsubsection*{Mixing dynamics, asymmetry, and minority size}
To uncover the regimes and scenarios of group inequality in the model, we characterize the impact of mixing and group sizes on average group degree. First, we study the trivial symmetrical mixing when $h_{00} = h_{11}$, and then we examine the impact of mixing asymmetry (i.e., $h_{00} \ne h_{11}$) on the social dynamics. Note that in asymmetrical case, the intragroup mixing are expressed by $h_{00}$ and $h_{11}$, whereas the intergroup mixing are the complementary probabilities, $h_{01} = 1- h_{00}$ and $h_{10} = 1 - h_{11}$.

\begin{figure}[t!]
\centering
\begin{adjustwidth}{-.2in}{0in} 
\includegraphics[width=6.48in]{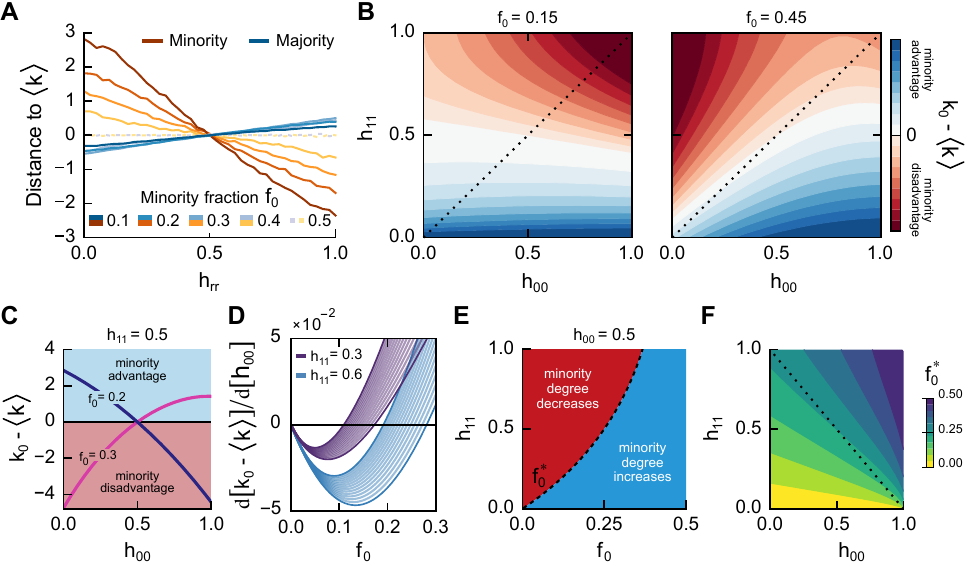}
\end{adjustwidth}
\caption{\textbf{The minority (dis)advantage in networks.} (\textbf{A}) The groups' average degree ($z$ score) with different minority fraction $f_0$ and at varying levels of symmetrical mixing (i.e., $h_{00} = h_{11} = h_{rr}$). The minority members have a degree advantage or disadvantage if the system is, respectively, a heterophilic ($h_{rr} < 0.5$) or homophilic ($h_{rr} > 0.5$) regime. (\textbf{B}) The distance of the minority group degree to the overall average degree, denoted $k_0 - \langle k\rangle$, at different levels of asymmetrical mixing (i.e., $h_{00} \ne h_{11}$). The majority mixing $h_{11}$ explains much of the variance of $k_0 - \langle k\rangle$. 
(\textbf{C}) The variation of $k_0 - \langle k\rangle$ with changes in the minority mixing $h_{00}$, with fixed $h_{11}=0.5$ and different minority fraction $f_0$. The minority mixing can have opposite impacts on degree inequality depending on the minority fraction, which suggests a qualitative transition in the system. (\textbf{D}) The derivative of $k_0 - \langle k\rangle$ as a function of $f_0$. The zero of this function represents the critical minority fraction, denoted by $f^\ast_0$, at which the qualitative transition occurs. In the plot, curves with the same color represent different values of $h_{00}$. (\textbf{E})~Two regimes delineated by $f^\ast_0$, with fixed $h_{00}=0.5$. These regimes mean that the minority group degree may either increase or decrease with a raise of $h_{00}$. (\textbf{F}) The parameter space of critical minority fraction. Given $h_{00}$, the upper limit of $f^\ast_0$, denoted as $\overline{f^\ast_0}$, represents the smallest minority size allowing higher minority homophily without decreasing group average degree, regardless of the majority mixing.
}
\label{fig:fig3}
\end{figure}

In the symmetrical case, we find that homophily and heterophily manifest themselves as two distinct regimes determining the position of the minority group in the network. We analyze the average degree of the groups given different values of group mixing $h_{rr}$ and minority fraction $f_0$. First, we simulate the model and measure the average degree of the minority $\langle k_0\rangle$ and majority $\langle k_1\rangle$ groups (see Methods). Then, we separately compare $\langle k_0\rangle$ and $\langle k_1\rangle$ to the average degree $\langle k\rangle$ of the whole network using their $z$ scores (Fig.~\ref{fig:fig3}A).  Our results show that the members of the minority group have an advantage or disadvantage depending on the model parameters.

Homophily ($h_{rr} > 0.5$) leads the minority group to be decoupled from a substantial part of the network. In this case, minority group members have a lower average degree compared to the average. In the heterophilic regime ($h_{rr} < 0.5$), the minority group has high visibility, which leads its members to have a higher degree than average. The existence of these two contrasting regimes implies that the very interpretation of a minority group depends on the group mixing in the network. When studying a minority group, one has to account for inter- and intragroup dynamics to understand its position in a network.

To characterize the impact of asymmetrical mixing on degree inequality, we examine the whole parameter space of the mixing matrix $\mathbf{H}$. We find that the majority mixing $h_{11}$ substantially contributes to degree inequality. First, we estimate the average minority degree for different values of $h_{00}$ and $h_{11}$, given specific minority fraction values $f_0$. Then, we measure the distance of the minority group degree to the overall average degree using  $z$ scores, denoted $k_0 - \langle k\rangle$ (see Fig.~\ref{fig:fig3}B). Our results show that the majority mixing $h_{11}$ explains much of the variance of $k_0 - \langle k\rangle$. While adjusting $h_{11}$ can change the position of the minorities from advantage to disadvantage, modifying $h_{00}$ can attenuate this inequality only slightly. 

\subsubsection*{Strategies for minority groups to alleviate degree inequality}

To uncover the ways the minority group can attenuate inequality, we investigate how $k_0 - \langle k\rangle$ varies with changes in the minority mixing $h_{00}$. We find that the size of the minority group modifies the system \textit{qualitatively}, revealing that changes in the minority mixing can have opposite impacts on degree inequality. For instance, given a constant value of $h_{11}=0.5$, increasing $h_{00}$ can reduce or accentuate degree inequality, depending on the minority fraction $f_0$ (see Fig.~\ref{fig:fig3}C). To characterize this  transition, we examine the derivative of $k_0 - \langle k\rangle$ with respect to $h_{00}$ as a function of $f_0$ (Fig.~\ref{fig:fig4}D). Precisely, we are interested in the zero of this function, which tells us the critical minority fraction $f^\ast_0$ at which the qualitative transition occurs. 
We find that this transition depends on the mixing dynamics of the system; its analytical form is given in Section 3.2.4 in the Supplementary Material. 
Because the exact analytical form is too intricate, we approximate it by assuming that the values of $h_{00}$ and $h_{11}$ are not at their extremes, finding that:
\begin{equation}
f^\ast_0 = \dfrac{h_{11}}{2(h_{01} + h_{11})}.
\label{eq:f0star_approximation}
\end{equation}
This equation demonstrates the control of the majority mixing over the minority group.
The critical minority fraction $f^\ast_0$ delineates two regimes where $k_0$ may either increase or decrease with a raise of $h_{00}$, given a fixed $h_{11}$ (see Fig.~\ref{fig:fig4}E). 

These regimes translate into two strategies to increase the minority average degree, which depends on the minority group size. First, when the minority group corresponds to more than $f^\ast_0$ of the network, increasing $h_{00}$ leads to an increase in the average minority degree. In this scenario, a more cohesive minority group is beneficial to its average degree. Second, in the case of a smaller minority group, increasing $h_{00}$ decreases its degree; a more cohesive minority comes with the cost of a less connected minority. In this case, decreasing $h_{00}$ helps in increasing the average degree of the minority group. 

To find when increasing minority homophily is always beneficial to the minority group, we characterize the upper limit of  $f^\ast_0$. This upper limit, denoted as $\overline{f^\ast_0}$, represents the smallest minority size in which the minority group can increase its homophily without detriment to its average degree, regardless of the majority mixing. We note that $\overline{f^\ast_0}$ is equivalent to the critical size $f^\ast_0$ when $h_{11}=1$ (see Fig.~\ref{fig:fig4}E), so its exact analytical form is:

\begin{equation}
\overline{f^\ast_0} =  \dfrac{1}{2 - h_{00} + \sqrt{3 + 2 (h_{00} - 2) h_{00}}}.
\end{equation}
%
%
For example, in the case of $h_{00}=0.5$, when the minority  represents more than 36.7\% of the population, the group increases its average degree by being more homophilic, regardless of how the majority group mixes.

\subsubsection*{Emergence of ranking misrepresentation}

\begin{figure}[b!]
\centering
\includegraphics[width=5.75in]{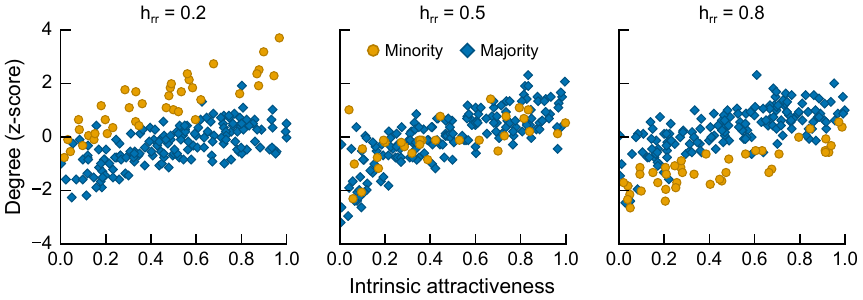}
\caption{\textbf{Same intrinsic attractiveness but different degree centrality. } 
The  intrinsic attractiveness of individuals versus their degree in different symmetrical regimes. In the neutral case ($h_{rr}=0.5$), individuals with higher intrinsic attractiveness have a higher degree regardless of their group membership. In nonneutral regimes, however, degree centrality disguises intrinsic attractiveness. In a heterophilic regime ($h_{rr}=0.2$), minority members have a higher degree compared to their majority counterparts with the same intrinsic attractiveness. In a homophilic regime ($h_{rr}=0.8$), highly attractive minority members do not have similar degree centrality to that of their majority counterparts. 
}
\label{fig:fig4}
\end{figure} 
From the model's viewpoint, mixing dynamics can inflate individuals' degrees because these dynamics tune individuals' intrinsic attributes. We show that group mixing can also lead social groups to be misrepresented in degree ranking. For instance, this misrepresentation is evident when we compare individuals' degrees with their intrinsic attractiveness in heterophilic, neutral, and  homophilic symmetrical regimes (see Fig.~\ref{fig:fig4}). Systematically, degree centrality disguises intrinsic attractiveness in nonneutral regimes. In a heterophilic scenario, minority members with low intrinsic attractiveness tend to have a higher degree than members of the majority group. In a homophilic situation, highly attractive minority members tend to have a lower degree than their majority counterparts. Because intrinsic attractiveness is hidden behind degree centrality, group members can be misrepresented in degree rankings.

\begin{figure}[b!]
\centering
\includegraphics[width=5in]{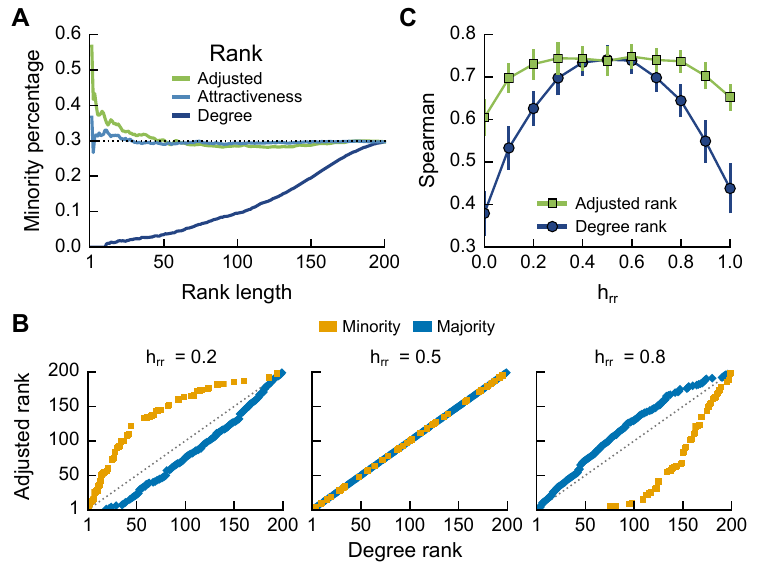}
\caption{\textbf{Adjusting for group mixing decreases misrepresentation in degree rankings.} (\textbf{A}) The percentage of minority members in rankings of different lengths (i.e., the top-\textit{k} rank). Each curve is a different ranking of a system with minority fraction $f_0=0.3$ and at a symmetrical homophilic mixing $h_{rr}=0.8$. The minority members are underrepresented in this regime; they are overrepresented in the heterophilic regime (see Section 3.2.6 in Supplementary Material). The adjusted ranking accounts for individuals' group membership, decreasing the misrepresentation of group members. (\textbf{B}) The adjusted ranking versus the degree ranking in different regimes. As expected, heterophily promotes minority members to higher positions in the degree ranking, whereas homophily pushes the minority down. (\textbf{C}) The Spearman correlation between the intrinsic attractiveness and the adjusted and degree ranking. The adjusted ranking tends to agree with the intrinsic attractiveness except in the extreme cases of $h_{rr} = 0$ and $h_{rr} = 1$.
}
\label{fig:fig5}
\end{figure} 

To characterize minority representation in rankings, we analyze individuals' rankings based on degree and intrinsic attractiveness. 
We rank individuals separately by degree and intrinsic attractiveness, and then we measure the minority percentage in each ranking as we increase the rank length (i.e., the top-\textit{k} rank; see Fig.~\ref{fig:fig5}A for a homophilic network). In the case of intrinsic attractiveness, we expect that the number of minority members in the top-\textit{k} rank is proportional to the minority size, since intrinsic attractiveness is uniformly distributed. This ranking displays this exact behavior (Fig.~\ref{fig:fig5}A). In the case of the degree ranking, however, the minority group has a substantially lower chance of appearing in top ranks than what we would expect from their attractiveness. These results indicate that degree ranking incorporates the mixing dynamics of the system (see Section 3.2.6 in Supplementary Material for a heterophilic case). This effect raises the question of how to decrease this misrepresentation. For example, one reason for adjusting rankings would be when ranking algorithms are used to rank and recommend people. In this case, despite its high intrinsic attractiveness, the minority's visibility will be disproportionately penalized in top ranks. 

\subsubsection*{Decreasing ranking misrepresentation}
To adjust the degree ranking, we need to account for the group sizes and mixing. Given that group mixing affects groups' average degree, we must compare only the degrees of individuals from the same group, thereby excluding the influence of group mixing. We show that this adjustment can balance degree rankings to contain a representative number of group members. First, we calculate the $z$ score degree of each individual with respect to their average group degree; then, we rank all individuals based on their $z$ score. Our results reveal that the adjusted degree ranking better represents the minority group, similar to what we would expect in the attractiveness ranking (Fig.~\ref{fig:fig5}A). The adjusted ranking also helps expose misrepresentation in degree rankings in different regimes. We compare individuals' positions in each ranking (Fig.~\ref{fig:fig5}B). As expected, heterophily promotes minority members to higher positions in the degree ranking, whereas homophily pushes the minority down. To characterize this misrepresentation, we analyze the correlation between the intrinsic attractiveness and the adjusted and degree rankings. We measure the Spearman correlation between the rankings and intrinsic attractiveness of nodes in different regimes (Fig.~\ref{fig:fig5}C). We find that the adjusted ranking tends to agree with attractiveness except in the extreme cases of $h_{rr} = 0$ and $h_{rr} = 1$. With the adjusted degree ranking, we decrease the misrepresentation bias in rankings. 

\section*{Discussion}


Face-to-face interaction is arguably a primary mechanism for the transmission and affirmation of culture\cite{Duncan1977}. When people interact with others, they construct a social world and participate in shaping the identity of social groups. In this work, we show that systemic degree inequality emerges in social gatherings from the way group members interact in imbalanced scenarios. Previous research has overlooked this inequality, suggesting that mainly the individuals' \emph{intrinsic attractiveness}  governs their connectivity. Our results indicate, however, that group dynamics modulate individuals' intrinsic attractiveness, forming what we have called social attractiveness in face-to-face situations.


In social gatherings, social attractiveness entangles with space and time variables, which restricts opportunities for face-to-face interaction, leading to systemic degree inequality. To interact, individuals must have opportunity availability (i.e., available place and time) and space--time convergence (i.e., individuals must agree on where and when to interact). In confined situations, such as conferences and workplaces, spatiotemporal constraints are critical to interaction opportunities. For example, there exists a limited number of opportunities for interaction at conferences (e.g., coffee breaks). When an individual uses an opportunity to interact with someone, fewer opportunities remain for interacting with other individuals. In imbalanced scenarios, when a majority member interacts with someone from the majority, fewer opportunities remain for this individual to interact with minorities, thereby decreasing minority connectivity. Such a decrease means that in the majority group position, creating homophilic ties comes at the cost of promoting inequality in group connectivity.


From a group-level perspective, mitigating this inequality depends primarily on the mixing of the majority group and its size---the minority group can only slightly reduce inequality, exhibiting a qualitative transition in the strategy for this reduction. Our results show that the majority group mixing explains most of the variance in connectivity inequality. To attenuate inequality, the minority group needs to follow a strategy that depends on its size. When the minority size is below a threshold in the network, homophilic minority interaction decreases minority connectivity. However, if the minority group size is sufficiently large, homophilic minority interaction helps in increasing minority connectivity. In this case, the minority group has proportionally enough individuals to interact with and decrease the disparity between groups. This critical mass allows the minority group to be a strong, tightly connected group; without this critical mass, a stronger minority implies higher inequality. This result is somewhat related to the critical mass for social change, as recently shown\cite{Centola2018}, in which committed groups with size higher than 25\% are sufficient to change social conventions. 

To summarize, we have investigated numerically, empirically, and analytically the emergence of group inequality in social gatherings. Our model allows us to study social interactions by accounting for mechanisms and spatiotemporal constraints in face-to-face situations. In contrast to previous works and non-mechanistic approaches (e.g., stochastic block models), our model captures the properties in face-to-face dynamics while reproducing degree inequalities found in data. The attractiveness--mixing model creates novel research opportunities to understand face-to-face situations. For example, the model can be used to explore how group mixing affects dynamic processes on social gatherings. Further research will help clarify the dynamic aspects of group inequality as well as latent groups in data. Likewise, the spatial aspect of social gatherings will be better understood when richer data containing spatial information is available. Moreover, though we have focused on a binary attribute, showing that social attractiveness can lead to group degree inequality, our model can also be used to investigate multiple and continuous-valued attributes. In the future, it would also be interesting to empirically estimate the model parameters such as the activation probability and intrinsic attractiveness, which would enable researchers to understand the role of these individual-level properties in social gatherings. 

In this article, we understand the concept of minority quantitatively: A minority group is the smallest group in a social gathering. Nevertheless, in the social sciences, the concept is often associated with the critique of inequality, deprivation, subordination, marginalization, and limited access to power and resources\cite{van1978minority}. 
Performing a quantitative analysis of the structural mechanisms of mixing, exclusion, and interaction by no means implies ignorance about these issues. This work sheds light on how inequality can emerge from social interaction, building new computational opportunities to understand and alleviate disparities in our society.


\section*{Methods}



\subsection*{Group mixing analysis}

To characterize how groups mix, we compare the inter- and intragroup edges in data with the configuration model. This approach enables us to assess whether the mixing patterns found in data would occur just by chance. With the configuration model, we generate random networks that preserve the degree of each node in a given network and reshuffle the links, which we can use to compare against the actual data\cite{maslov2002specificity}. For each data set, first, we generate $500$ random instances of the network and count the number of edges $E'_{rs}$ between groups $r$ and $s$ in each instance; then, we count the actual number of edges $E_{rs}$ in the data; finally, we compare $E_{rs}$ to $E'_{rs}$ via $z$ scores, defined as $z_{sr} = ({E_{sr} - \overline{E'}_{sr}})/{s[E'_{sr}]}$, where $\overline{E'}_{sr}$ and $s[E'_{sr}]$ are, respectively, the mean and standard deviation of ${E'}_{sr}$ over the $500$ instances. The $z$ score $z_{sr}$ reveals the number of standard deviations by which $E_{rs}$ differs from the random case.

\subsection*{Model and  simulation analysis}

We summarize the main entities of the model and paper in Table~\ref{tablevariables}. In the simulations, we used the following parameters: $L=100$, $d=1$, $v=1$, and $N=200$. To analyze the symmetrical case of the attractiveness--mixing model, we first simulate the model with different values of $h$ and $f_0$ and then measure the average degree of the whole network $\langle k\rangle$ and the average group degree of the minority, denoted as $\langle k_0\rangle$, and majority, $\langle k_1\rangle$. We compare each group $r$ with the whole network using $z$ scores, defined as $(\langle k_r\rangle - \langle k\rangle)/{s[k]}$, where $s[k]$ is the standard deviation of $k$. 
In the case of the simulations using the data estimates, we run the model until the total number of edges is the same as the number of edges in the data.
In all analyses, we average the results over $50$ different simulations.

\begin{table}[h]
\caption{Variables, constants, and notations of the model.\label{tablevariables}}
\centering
\begin{tabular}{cl}
\toprule
Entity & Description   \\\midrule
$E$ & Number of edges in a network (or system) \\
$N$ & Number of nodes in a network (or system) \\
$B$ & Number of node groups in a network (or system) \\
$n_r$ & Number of nodes in group $r$  \\
$f_r$ & Group fraction, defined as $n_r/N$  \\
$E_{rs}$ & Number of edges between group $r$ and $s$  \\
$e_{rs}$ & Normalized edge count, defined as $E_{rs}/E$   \\
\midrule
$\mathbf{H}$ & The mixing matrix  \\
$v$ & Step length  \\
$L$ & Space size  \\
$d$ & Neighborhood radius  \\
$b_i$ & The group label of the individual $i$  \\
\midrule
$r_i$ & Individual $i$'s activation probability  \\
$\eta_i$ & Individual $i$'s attractiveness  \\
\bottomrule 
\end{tabular}
\end{table}

\subsection*{Analytic derivation of the model}
Here we derive the attractiveness--mixing model analytically for the case of two groups, $B=2$, denoted as group $0$ and group $1$. We show that we can calculate the normalized group edge matrix analytically. To simplify the derivation, we assume a situation of low spatial density of agents; thus, almost all situations of potential interaction involve only two individuals, and non-pairwise interactions are negligible (i.e., dilute system hypothesis). Due to the activation process, the average number of active nodes at one time $t$ is $N_a = \left<r_i\right>N$, since we do not expect a correlation between $r_i$ and the presence of isolated individuals. Therefore, the number of active nodes in each group is $N_0 = f_0N_a$ and $N_1 = f_1N_a$. We note that the attractiveness--mixing model generates a temporal network in which nodes are the individuals and interactions between them are edges that are created and destroyed as time passes. Let $E$ be the number of edges created during a time step $\Delta t$. Since the interaction mechanism is time-independent, the total number of edges in the network after a time $T$ is given by $E_T = E \times {T}/{\Delta t}$.

To express the number of new edges created at time $t$ in each component, first, we focus on edges between two individuals from group $0$. The probability $p_0$ to find an individual $j$ from group $0$ in the vicinity of individual $i$ relates to the surface of the vicinity and the density $\rho_0$ of individuals from group $0$ on the field, expressed by: 
\begin{equation}
p_0 = \rho_0S = \dfrac{N_0}{L^2} \pi d^2 = \dfrac{f_0N_a}{L^2}\pi d^2 = f_0\rho_a\pi d^2,     
\end{equation}
 where $\rho_a$ is the density of active individuals. To have an interaction, \emph{both} individuals need to be available and not move away. In our case, the probability for one individual to be available is the attractiveness of the other individual; thus, the probability of having both is the product $\eta_i\eta_j$, whose average value is a constant $\gamma  = \left<\eta_i\eta_j\right>$.
    %
%
We note that three situations can lead to the creation of an edge between an individual $i$ and an individual $j$: (1) only individual $i$ initiates the creation, (2) only individual $j$ initiates the creation, or (3) both individuals initiates the creation.
Therefore, 
the probability for the edge to appear is thus:
\begin{equation}
p_{ij} = p_{i\to j} + p_{j\to i} + p_{i\leftrightarrow j}.    
\end{equation}
In the case of two individuals from the group $0$, this probability is on average:
\begin{align*}
  p_{ij,00} &= f_0\rho_a\pi d^2 \times \gamma \times \big[h_{00}(1-h_{00}) + (1-h_{00})h_{00} + h_{00}^2\big]\\
         &= f_0\rho_a\pi d^2\gamma\big(1 - h_{01}^2\big). \addtocounter{equation}{1}\tag{\theequation}
\end{align*}
%
Thus, the number of intra-group edges created during one time step for group $0$ is:
\begin{equation}
E_{00} =  \dfrac{N_a^2\pi d^2}{2L^2}\gamma f_0^2\big(1 - h_{01}^2\big),
\end{equation}
where the $1/2$ factor takes into account double counting. With a similar approach, we find that
\begin{equation}
E_{11} = \dfrac{N_a^2\pi d^2}{2L^2}\gamma f_1^2\big(1 - h_{10}^2\big).
\end{equation}
Furthermore, this procedure also enables us to find that
\begin{equation}
E_{10} = \dfrac{N_a^2\pi d^2}{L^2}\gamma f_1f_0\big(1 - h_{00}h_{11}\big)
\end{equation}
and
\begin{equation}
E_{01} = \dfrac{N_a^2\pi d^2}{L^2}\gamma f_0f_1\big(1 - h_{00}h_{11}\big), 
\end{equation}
where the $1/2$ factor is not required in these cases because they do not have double-counting. Finally, since $E=E_{00} + E_{01} + E_{10} + E_{11}$ by definition, the share of intra-group edges for group $0$ is given by:
\begin{equation}
e_{00} = \dfrac{E_{00}}{E}=\dfrac{f_0^2\big(1 - h_{01}^2\big)}{f_0^2\big(1 - h_{01}^2\big)  + 2f_0f_1\big(1 - h_{00}h_{11}\big)+ f_1^2\big(1 - h_{10}^2\big)}.
\label{eq:e00}
\end{equation}
By following analogous procedure, we can find $e_{11}$ as written in Eq.~(\ref{eq:e11}). In addition, $e_{01}$ and $e_{10}$ can be written as
%
\begin{equation}
e_{10} = e_{01} = \dfrac{f_0f_1\big(1 - h_{00}h_{11}\big)}{f_0^2\big(1 - h_{01}^2\big) + 2f_0f_1\big(1 - h_{00}h_{11}\big) + f_1^2\big(1 - h_{10}^2\big) }.
\label{eq:e10}
\end{equation}
 We verify that these equations predict the model behavior well by comparing them with  simulations (see Fig.~5 and Fig.~6 in Supplementary Material). Furthermore, we use Eq.~(\ref{eq:e11}), Eq.~(\ref{eq:e00}), and Eq.~(\ref{eq:e10}) to estimate $h_{00}$ and $h_{11}$ from data via numerical optimization. First, we calculate the empirical share of intra- and inter-group edges in the networks, then we use an optimization algorithm to find the corresponding values for $h_{00}$ and $h_{11}$.

\bibliography{sciadvbib}
\bibliographystyle{naturemag}

\noindent \textbf{Acknowledgments:} 
We thank Lisette Esp\'{i}n-Noboa, Haiko Lietz, Diego Pinheiro, and Diogo Pacheco for helpful feedback on the manuscript. 
\textbf{Competing interests:}
 The author declare no competing interests.

\end{document}